\documentclass[12pt,preprint,round,colon,authoryear,longnamesfirst]{aastex}

\begin{document}

\shortauthors{Metchev et al.}
\shorttitle{Mass Limits on Companions to Vega}

\title{Adaptive Optics Observations of Vega: Eight Detected Sources and
  Upper Limits to Planetary-Mass Companions}
\author{Stanimir A.\ Metchev, Lynne A.\ Hillenbrand, Russel J.\ White}
\affil{California Institute of Technology}
\affil{Department of Astronomy, MC 105-24, Pasadena, CA 91125}
\email{metchev, lah, rjw@astro.caltech.edu} 

\received{}
\revised{}
\accepted{}
\journalid{}{}
\articleid{}{}

%\clearpage

\begin{abstract}
From adaptive optics observations with the Palomar 5-meter telescope we
place upper limits on the masses of any planetary 
companions located between $\sim$30--230~AU away from Vega, where our
data are sensitive to depths ranging from $H=12.5$~mag to $H=19.0$~mag
fainter than Vega itself.  Our observations cover a plus-shaped area
with two $25\arcsec \times 57\arcsec$ elements, excluding $7\arcsec
\times 7\arcsec$ centered on the star.   We have identified 2 double
and 4 single point sources.  These projected companions are
14.9--18.9~mag fainter than Vega, and if physically associated would 
have masses ranging from 4 to 35~M$_{\rm Jup}$ and orbital radii
170--260~AU.  Recent simulations of dusty rings around Vega  predict the
presence of a perturbing body with mass $<$2--3~M$_{\rm Jup}$ and
orbital radius $\sim$40--100~AU, though more massive ($\lesssim$10~M$_{\rm
Jup}$) planets cannot be excluded.  None of the detected objects are this 
predicted planet.  Based on a color-magnitude, spectroscopic, and proper
motion analysis, all objects are consistent with being background sources.
Given the glare of Vega, a 2~M$_{\rm Jup}$ object near the expected
orbital radii would not have been visible at the 5$\sigma$ level in our
data, though any $>$10~M$_{\rm Jup}$ brown dwarf could have been seen at
separation $>$80~AU.
\end{abstract}


%\clearpage

\section{INTRODUCTION \label{sec:intro}}

The A0V star Vega is famously known since the early days of data return
from IRAS as a young main sequence star surrounded by dust \citep{aum84}.
Its age \citep[270--380~Myr;][]{son01} combined with the
large fractional excess luminosity at infrared wavelengths \citep[$L_{\rm
excess}/L_\ast \approx 10^{-5}$ or $M_{\rm dust} \approx 1/2 M_{\rm
moon}$;][]{bac93} imply that dust is being generated at the current
epoch by either grinding collisions between larger rocky bodies, a.k.a.\
planetesimals \citep*{harp84,wei84,zuc93}, or in cometary ejecta
\citep[and references therein]{beu89,beu90}. If the dust is not
continuously regenerated it will be depleted by a 
combination of Poynting-Robertson drag and radiation pressure
on a time-scale much shorter than the age of Vega.
Discovery of the infrared excess around Vega and other main sequence stars too
old to possess the so-called primordial dust and gas disks that are commonly
found around 1--10~Myr old stars, led to coining of the term ``debris disk.''
Searches for other examples of ``the Vega phenomenon'' have led to
cataloging of a mere tens of objects \citep[see, e.g.][]{man98,sil00},
mostly early-type stars whose dust was detectable with {\sl IRAS} or
{\sl ISO}, or observable from the ground with mid-infrared
instrumentation on large telescopes.
  
The mid- and far-infrared (25--850~$\micron$)
emission from Vega is extended
over tens of arcseconds \citep*{aum84,harv84,zuc93,hei98,hol98}.
Aperture synthesis imaging at 1.3~mm \citep*{koe01,wil02}
resolved several dust clumps located $\sim$8--14$\arcsec$ from the
central source (60--110~AU, assuming the Hipparcos parallax of
128.9~milli-arcsec).  One interpretation is that these clumps trace
the densest portions of the already inferred face-on circumstellar ring 
\citep{den00}.  Additional support for a ring interpretation comes from
Vega's spectral energy distribution, which is close to photospheric at
shorter wavelengths \citep[$\lesssim$20~$\micron$;][]{hei98}, and
suggests an inner gap in the density distribution which may or may not
be entirely devoid of hot dust.  At 11.6~$\micron$ extensions larger
than 1/4$\arcsec$ are ruled out by the imaging of \citet*{kuc98}.
Interferometric work by \citet{cia01}, 
however, did suggest extended emission at 2.2~$\micron$. 

Observations of structure in the circumstellar
dust around Vega have spawned detailed models for a planetary perturber
\citep{gor01,wil02}. Resonance trapping and gravitational scattering
induced by a body of mass 2--3~$M_{\rm Jup}$ are 
consistent with the \citet{hol98} map, and 
with the interferometric observations of \citet{koe01} and
\citet{wil02}.  Due to degeneracies in dynamical models
\citep[e.g.,][]{wil02}, more massive planets ($\sim$10~M$_{\rm Jup}$)
also cannot be ruled out.  Modeling to date assumes a face-on
orientation of the  
presumed dust disk or ring.  Evidence for this geometry comes both from
a ring-shaped \citep[e.g.,][]{hei98} albeit clumpy \citep{koe01,wil02}
dust distribution, and from detailed analysis of stellar line profiles
\citep*[assuming parallel disk and stellar rotation axes;][]{gul94}.

Our experiment was designed to search for low-mass companions within
4--30$\arcsec$ of Vega, in part to test the aforementioned planetary
perturber predictions.  Imaging observations close to this bright source
are usually ``burned out'' in survey data such as POSS or 2MASS.
Ground-based coronagraphic observations \citep*{smi92,kal96} have also
lacked sufficient sensitivity. Except for NICMOS images \citep*{sil02}
with sensitivity comparable to ours, high dynamic-range observations
have not been previously reported.

\section{OBSERVATIONS \label{sec:obs}}

Data were obtained with the Palomar adaptive optics
\citep[PALAO;][]{tro00,blo00} system in residence at the
Palomar 5-m telescope.  PALAO employs PHARO, the Palomar High Angular
Resolution Observer \citep{hay01}, a 1024$^2$~pix HgCdTe HAWAII detector
with imaging ($25\arcsec$ or $40\arcsec$ field of view) and
spectroscopic ($R=$1500--2500) capabilities.  Broad- and narrow-band
filters throughout the $JHK$ atmospheric windows are available, as well
as a choice of coronagraphic spot sizes and Lyot masks.

Vega was observed on the night of 2002 June 22 with additional
follow-up observations obtained on August 28 and 29 (UT), all under
photometric sky conditions.  The observing strategy 
was to take deep images in $H$-band to maximize
the detection likelihood of faint low-mass objects \citep[see
e.g.,][]{bur97}.  The point spread function
(PSF) was 0.6--0.9$\arcsec$ uncorrected at $H$-band, and improved to
$<$0.1$\arcsec$  with adaptive correction.  A neutral density
filter (1\%) manually placed in front of the wave front sensor (WFS)
enabled AO lock on such a bright object.  AO performance was very good
during most of the observing, with Strehl ratios up to 20\% in
$H$.  We did not employ the coronagraphic mode of PALAO for 
these observations since scattered light suppression was not sufficient
enough to prevent saturation on the array outside the boundaries of the
largest coronagraph ($0.97\arcsec = 12\lambda/D$ in $H$) in the shortest
possible integration time (1897~milli-sec).   

On June 22, a total of 26 minutes on-source integration was obtained
with the $25\arcsec$ field of view in $H$-band at each of 4 pointings:
north, south, east, and west around Vega (henceforth: Vega N, S, E, and
W fields), with Vega itself located $3.5\arcsec$ off of the imaging
field at each positioning of the telescope.  Due to field overlaps,
$\sim$13\% of the area covered (2210~arcsec$^2$) was observed for 52
minutes. Dithering at the 0.25--1.00$\arcsec$ level was 
performed for the on-source frames. More widely dithered sky frames were
taken at locations $\sim$2$\arcmin$ further away from Vega with
source-to-sky time split 2:1. The integration time for individual
exposures was 10.9~seconds.  For the eastern field in
which several objects were noticed in real time, we also obtained
$J$, $H$, and $K_s$ data with 2.5 minutes total
on-source integration time taken as 5 separate frames, with Vega offset
22--28$\arcsec$ to the west.  The airmass range was 1.03--1.30
for the entire observing sequence.

Photometric calibration was achieved via immediate observation of 2MASS
183726.28+385210.1 (GSC 03105-00679, a G8V star) 
located $\sim$7.7$\arcmin$ from Vega with 2MASS magnitudes 
$K_s=8.296 \pm 0.033$, $H=8.365 \pm 0.022$, and $J=8.745 \pm
0.028$. This source, although not a photometric standard, is
sufficient as a local calibrator and was observed at airmass 1.35.  Two
other much fainter 2MASS sources are also present in the image. 

During the second epoch observations, resolution $R=1500$ and 2400
$K$-band spectra of the brightest discovered object were obtained (August 28)
through a $0.52\arcsec$ slit and a $K$ grism for a total of 100~minutes
on-source integration.  The object was dithered $10\arcsec$ along the
slit for sky-subtraction.  Spectra of scattered light from Vega were
used as a telluric standard.  Short-exposure (5 minutes per filter)
dithered $JHK_s$ images were taken (August 29) as follow-up to the June
22 data to test for common proper motion with Vega.  The airmass of Vega
for the second epoch observations varied between 1.01 and 1.13.

We also observed a binary system (HD~165341) with 
a well-determined orbit in the Sixth Catalog of Orbits of Visual 
Binary Stars\footnote{available at http://ad.usno.navy.mil/wds/orb6.html} 
in order to determine precisely the plate-scale and orientation of the PALAO 
system.  We derive for the $25\arcsec$ field a plate-scale of $0.025168 \pm
0.000034$~arcsec/pixel.

\section{DATA PROCESSING}

Our image reduction steps, written in IDL and IRAF, include the standard
procedures of flat-fielding, sky-subtracting, interpolating/masking bad
pixels, and mosaicking the dither pattern.  This last step
required correcting for image drift (likely caused by change in the
direction of the gravity vector in PHARO over the duration of the
observing sequence), the rate of which varied between 0.5 and
1.2~arcsec/hour. 

Image stacks from each of the four deep pointings (June 22) were registered
to the first image in the series. For the east field in which several
point sources were detected, each image was registered by centroiding on
the brightest object.  For the north field, centroiding was possible on
the bright reflection artifact due to Vega. For the other two fields
registration was accomplished by first averaging sets of 9 consecutive
exposures, extrapolating the  position of Vega from the intersection of
6 scattered light ``rays'' in the image halo, and combining the 16 registered
averages.  In this manner, the location of the star could be constrained
to within $\pm 5.0$~pix $=\pm 0.13\arcsec$ (c.f.\ $\pm 0.10$~pix for our
mean centroiding precision in the north and east fields).  We did
attempt cross-correlation techniques for dither pattern correction but
these were not as successful as the above procedures.  The final step
was to orient the images with north--up and east--left. Astrometric
calibration was established assuming the plate-scale derived from the
binary star observations and  by reference to the Hipparcos (J2000.0)
coordinates of Vega. Our final image of the Vega vicinity is presented
in Figure~\ref{fig_pharo}a.

Various methods to reduce the large halo from Vega were attempted, including
reflections, rotations, and data smoothing.  Shown in Figure~\ref{fig_pharo}b
is a difference image, for which a Gaussian-smoothed ($\sigma=5$~pix, ${\rm
FWHM}=12$~pix; c.f.\ ${\rm FWHM}=4$~pix for the point sources) version
of the original image has been subtracted.  This procedure effectively
removes large scale gradients.  Strong artifacts do remain, however, and
contribute to our limited sensitivity to point sources within
$\sim$10$\arcsec$ of Vega.

Spectra of the brightest point source were extracted using the APALL task 
within IRAF.  A quadratic polynomial was fit to all pixels with values
$>$10\% of the peak flux along an aperture.  Local background was
estimated from a region 0.50--1.25$\arcsec$ from the aperture center.
The extracted spectra were divided by that of the telluric
standard (with the 2.166~$\micron$ Br$\gamma$ absorption feature
interpolated over) to correct for instrumental response and atmospheric
transmission.  Wavelength calibration was done by fitting a dispersion
relation to sky OH emission lines.  Finally, the wavelength-calibrated
spectra were co-added.

\section{PHOTOMETRY OF DETECTED SOURCES \label{sec_phot}}

The positions of identified sources are indicated in
Figure~\ref{fig_pharo}.  During the first epoch of observations we
detected 6 point sources, 5 of which were
to the east of Vega.  Four of these are single 
while two are a close ($0.6\arcsec$) double, which is $0.7\arcsec$
off the east edge of our deep Vega E image, and was observed only in the
short $JHK_s$ exposures.  The sixth point source is in the north field.
During the follow-up observations we detected another double
($0.8\arcsec$) source south-east of Vega.
PSF fitting techniques suggest these are in fact all stellar point
sources and not partially resolved galaxies.

We performed photometry using both aperture and PSF techniques.
First, we used the IRAF/PHOT task in the short exposures with
aperture radii of 10 ($K_s$), 18 ($H$), and 32 ($J$) pixels
($0.50\arcsec, 0.90\arcsec$, and $1.5\arcsec$ diameters on the sky)
chosen to correspond to $2\times {\rm FWHM}$ of the image core, and to
include the first Airy ring. The
mode of the counts in a 30--40 pixel annulus provided local sky
which was critical for subtracting residual scattered light from Vega.  
For sources 1--6, magnitudes in each of the bands were obtained by
comparing the measured aperture flux to that of the 2MASS standard in
the same aperture. The magnitudes and positions of sources 7 and 8 were
boot-strapped from those of source 1, with its error added in quadrature.
Airmass corrections were applied using extinction coefficients for
Palomar as previously determined by L.A.H.\ (0.114, 0.029, and 0.065
magnitudes per airmass in $J$, $H$, and $K$, respectively).
We also used the PSF, PEAK, and ALLSTAR tasks in IRAF/DAOPHOT
for PSF fitting photometry.  PSF fitting worked best at
$K_s$-band but required a large number of iterations at $H$ and $J$ for
convergence in part because the stellar profiles are not diffraction
limited.  Differences  between the aperture and PSF-fitting magnitudes
are 0.2--0.3~mag (much larger than the formal 
errors), and the scatter of the PSF magnitudes is 50\% larger than that
of the aperture magnitudes at $J$ and $H$. 

Our photometry (Table~\ref{tab_phot}) is
from apertures, except for sources 4 and 5, for which we simultaneously
fit PSF profiles to each of the components of the double source to
determine their magnitude difference.  A larger aperture ($2.5\arcsec$
diameter~-- to include the PSFs of both sources at all bands) is
used to measure a combined flux, and individual magnitudes are 
obtained from the large-aperture magnitude and the magnitude difference from
PSF fitting.  The photometry for these two sources is less precise due
to a more uneven background. 

Repeatability of the photometry from frame to frame was assessed using
aperture photometry on the calibration field, which is free of the bright
background present in the Vega fields.  We find 0.04~mag r.m.s.\ scatter
between the 5 frames.  For the shallow $JHK_s$ exposures
near Vega, frame-to-frame differences are larger due to background
variations induced by dithering which placed Vega closer to the image
area for some frames than for others.  We have included this scatter in
our errors.  

We do not include a Strehl term in our calibration, as the implied
corrections were larger than the uncorrected frame-to-frame scatter.  The
Strehl ratio changed from $\sim$15\% in the deep $H$ exposures to 2--3\%
in the subsequent shallow ones, but was relatively stable between the
short exposures of the object and the calibration fields.

\section{ANALYSIS}

\subsection{Sensitivity Limits}

In the absence of the bright glare from Vega, our deep observations
should nominally detect point sources at $S/N=5$ to $H=20.8$ (21.2, for
13\% of the image), while the shorter $JHK_s$ exposures should  
reach $J=20.8$, $H=20.1$, and $K_s=18.9$.  However, the star adds
substantial scattered light background and makes point source detection
a function of position with respect to Vega.

We have assessed our $H$-band detection limits using artificial star
experiments, both in the direct image mosaic, and in the halo-subtracted
image.  IRAF/PSF was used to fit the two brightest single objects in
the processed Vega E image, and artificial stars were added to
the same image with ADDSTAR.  A single experiment consisted of
adding sources of constant magnitude at $1\arcsec$ intervals along 9
radial (originating from Vega) directions, offset by $15\degr$ from each
other. We observed the minimum separation from Vega, at which a source
would be considered ``detected'' by eye: at $S/N\gtrsim5$ according to
formal $S/N$ calculations assuming Gaussian noise statistics.  
Since the primary source of noise (scattered light
form Vega) does not behave in a Gaussian manner however, the $S/N$ statistic
does not carry the correct information about the significance of a 
detection, and is only used as an approximate measure of the local contrast.

The experiments were repeated at 0.5~mag steps.  For a given radial
distance, there are thus up to 9 independent measurements of the
limiting magnitude (fewer for larger distances, where some artificial
sources fall beyond the array), as shown in Figure~\ref{fig_sens}.  Our
average sensitivity ranges from $\Delta H=12.5$~mag at $4\arcsec$ to
$\Delta H=19$~mag at $\geq$26$\arcsec$, 1.8 magnitudes brighter than for
low-background observations.

Extensive artificial star experiments were not performed for the less
well-registered parts of the mosaic: the Vega S and Vega W fields.
However, after applying the registration method used for these fields (ray
intersection) to the Vega E field, for which centroiding provided
the best registration among our fields, we observe that the
faintest object in Vega E ($H=18.9$) is at the detection limit
($S/N=4.7$).  The detection limit is thus 0.3~mag brighter than
the $H=19.2$ found at that location using centroiding.  Since smearing
of point sources due to improper registering is uniform across the image
(there being only translational and no rotational degrees of freedom),
we estimate limiting magnitudes in Vega S and W $\sim$0.3~mag brighter
than in Vega E and N.

\subsection{Comparisons to Models}

\begin{figure}
\plotone{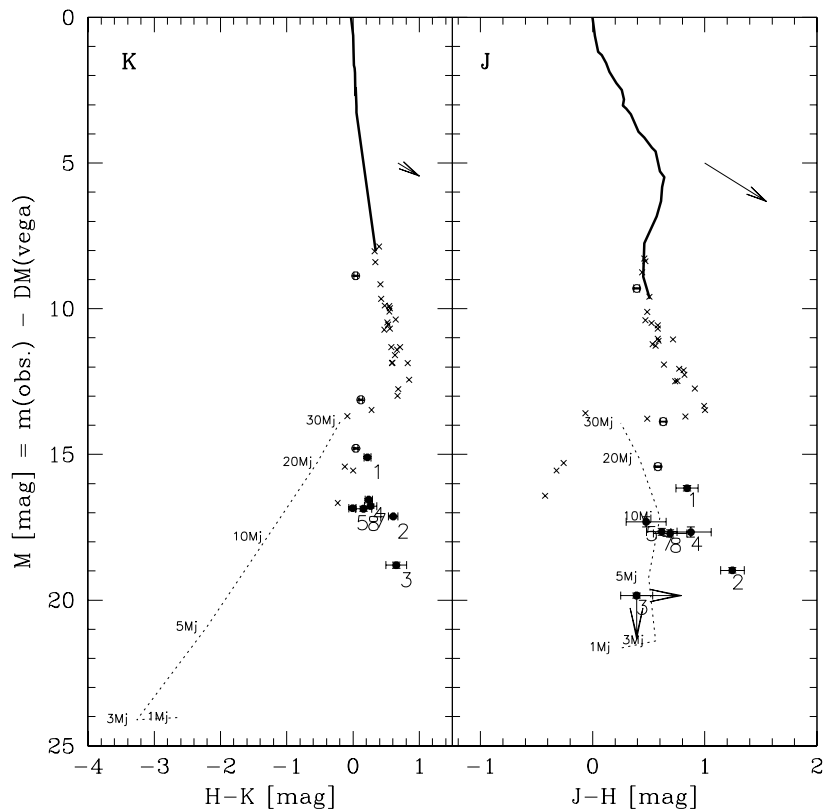}
\figcaption{$JHK_s$ color-magnitude diagrams, in the CIT photometric
  system. Heavy solid line is 
  the main sequence relation for spectral types A0--M6 and the
  crosses are M4--T6 dwarfs from \citet{leg02}.  Dotted line
  is the \citeauthor{bur01} 300~Myr isochrone for masses 1--30~$M_{\rm
  Jup}$, as labeled.  The arrow corresponds to 5 magnitudes of
  interstellar reddening.  Filled circles with error bars represent
  our Vega field data, while open circles are the calibration field
  data.
\label{fig_comb}}
\end{figure}

\begin{figure}
\plotone{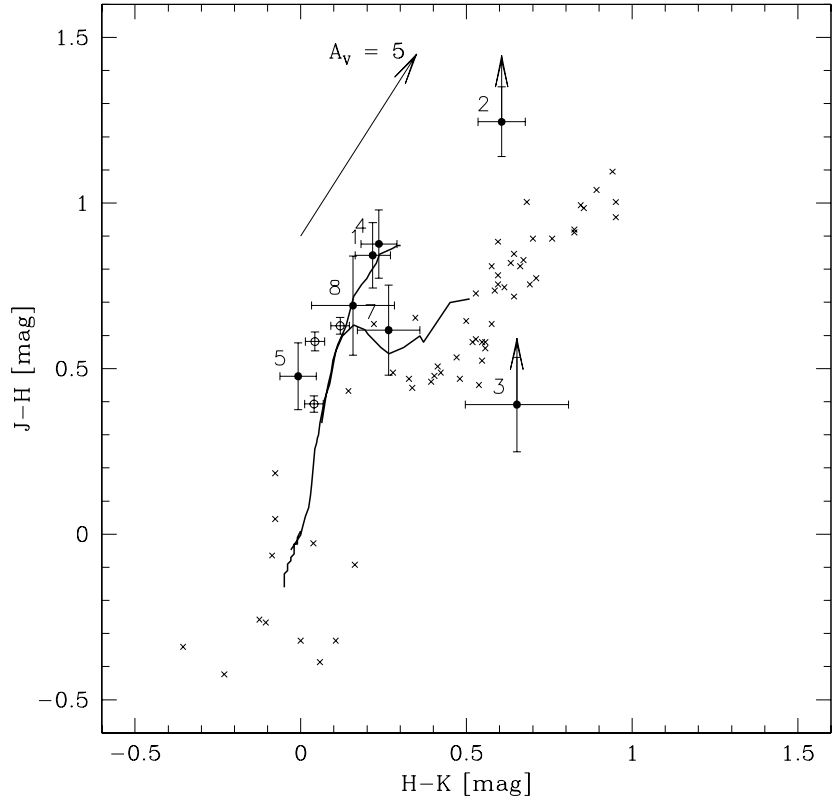}
\figcaption{$JHK_s$ color-color diagram, in the CIT photometric system.
  See Figure~\ref{fig_comb} for description of symbols.
\label{fig_jhhk}}
\end{figure}

Figures~\ref{fig_comb} and \ref{fig_jhhk} show the
photometric measurements from Table~\ref{tab_phot} for the detected
point sources assuming a common distance modulus with Vega,
along with a 300~Myr isochrone for 1--30~$M_{\rm Jup}$
objects \citep{bur01}, and known field L and T dwarfs (whose ages may
range from 0.5--10~Gyr).  Given their colors, all sources detected by us
in the vicinity of Vega are too red
compared to the expected locus of planetary-mass companions (from
Table~\ref{tab_phot}) and too faint to be brown dwarfs.  Hence, they are
most likely background stars.  This was confirmed for objects 1--5 by a
relative proper motion test, with the positions of objects 2--5
(measured from PSF fits) compared to that of object 1.  None
changed by more than $17\pm15$~milli-arcseconds (mas) =
$0.68\pm0.60$~pix between the  
two epochs.  The proper motion for Vega over the period (67 days) was
$64.4\pm0.8$~mas = $2.56\pm0.03$~pix (Hipparcos), and hence any
projected companion that is gravitationally bound to Vega should have
moved by this amount (barring all 5 being associated).

No colors or proper motion information are available for source 6,
hence we can only estimate its likelihood of association with Vega from
the expected frequency of field stars.  To assess background
contamination, we used the SKY model of \citet{wai92}, which for the
position of Vega ($l=67.45\arcdeg, b=19.24\arcdeg$) gives a 7.6\%
probability that 4 or more stars of the specified magnitudes (for
objects 1--3 and 6) are seen in the deep image.  Thus, our detections are
statistically consistent with being background stars.

Our results, nevertheless, demonstrate that detection of planetary-mass
companions to nearby stars with ground-based telescopes is a real
possibility. Based on their $H$ magnitudes, we list the predicted masses
of the candidate companions in the last column of Table~\ref{tab_phot},
assuming a common distance modulus with Vega, and using models
from Burrows (private communication) and \citet{cha00} for a 300~Myr old
star.  In using the \citeauthor{cha00} models, linear interpolation has
been applied between the values for 100~Myr and 500~Myr.
Both sets of models include internal heating processes only and not, e.g.\
irradiation of the planetary atmosphere by the star or reflected light
from the star, but are appropriate given the large orbital separation of
our candidate companions and the wavelength regime in which we are
working.   We should have detected any planets/brown dwarfs
$>$10~$M_{\rm Jup}$ at separations $>$12$\arcsec$ (90~AU), and
$>$5~$M_{\rm Jup}$ at $>$20$\arcsec$ (160~AU) from Vega.

\section{DISCUSSION}

Based on proper motion, colors, and field star considerations, it is
unlikely that the newly discovered objects are sub-stellar companions to
Vega.  Yet their existence in close proximity to Vega is heretofore
unappreciated. 
 
With respect to the predicted 2--3~$M_{\rm Jup}$ planetary perturber to
Vega's dust 
distribution, \citet[see also \citet{oze00}]{gor01} quote a exact
positions of a possible planet, with orbital radius $\sim$100~AU.
We find no $H<16.5$ objects ($>$8~$M_{\rm Jup}$; \citeauthor{bur01}
models) at either of their quoted positions, or along the line connecting
them, which may also be solutions to the model.  Point sources are found
neither along the \citeauthor{wil02} planetary orbit, nor anywhere within
the $14\arcsec$ sub-millimeter emission (albeit at lower sensitivity
limits: $H<17$--13; $>$7--30~$M_{\rm Jup}$).  

How do our upper limits compare to others in the literature for Vega?
\citet{gat95} find no astrometric evidence for planets $>$12~$M_{\rm Jup}$
at 1.5--5.0~AU (1.2--7 year period).  \citet{hol98} place an
upper limit of 12~$M_{\rm Jup}$ on companions based on  
null result observations with Keck/NIRC, though no details are given.
The NICMOS images of \citet{sil02} have similar sensitivity to ours (to
within 0.5~mag at 1.10~$\micron$ and 2.05~$\micron$), however cover 
an area too small to see any of the objects detected by
us. The \citet{opp99} survey of stars within 8~pc, which just barely 
included Vega, found no companions at the positions of our detections.  
Based on their sensitivity curves, objects brighter than $r=$16--17~mag
should have been detected around Vega from 20--30$\arcsec$.   Given the
$R-H$ colors of low-mass stars for which $H-K=0.1$--0.3 (K2--M5 spectral
types: $R-H>$2.0), the \citeauthor{opp99} survey may have just missed
detecting our brightest projected companion to Vega if it is a
background star as early as K2.  Our spectrum ($S/N\approx15$) of object
1 indeed places it in the K5V--M5V spectral type range.

Our imaging data can also be used to test a possible cosmological origin
of the sub-millimeter dust clumps around Vega.  Spectral energy
distributions of (sub-)millimeter galaxies \citep{dan02,gol02,kla01}
suggest $z\gtrsim$1 for any responsible background galaxy, given our
non-detection at $H$ band.  However, deep searches for $K$-band
counterparts to several sub-millimeter galaxies have reached $K\approx
22$~mag \citep[e.g.,][]{dan02} with no counterpart detection, suggesting
that our data may be too insensitive (by 6--7 magnitudes at these
locations) to put a sensible limit on this hypothesis.

\section{CONCLUSIONS}

We find 8 faint objects within $35\arcsec$ of Vega that are 15--19~mag
fainter than the star at $H$-band. If associated, at the 330~Myr age for
Vega, current brown-dwarf cooling models \citep{bur01,cha00} set their masses
at 5--35~$M_{\rm Jup}$.  The number of detected objects is however
consistent with estimates of field star density, and their colors and
proper motion indicate that they are not associated with Vega.

We thus exclude the possibility of a distant (80--220~AU; $\sim$83\% of
this area is imaged), massive ($>$10~$M_{\rm Jup}$; $>$6~$M_{\rm Jup}$
for 120--220~AU) planetary/brown-dwarf companion causing the 
observed dust distribution around Vega. We also detect nothing at the
positions of the predicted planetary perturbers, with upper mass limits
7--15~$M_{\rm Jup}$ ($H$$<$$17$--13), well above the 2--3~$M_{\rm Jup}$
predictions.  We detect nothing at the position of the mid-infrared dust
clumps, placing limits on the possibility of their extragalactic
interpretation.

\begin{acknowledgements}
We acknowledge with appreciation those who have endeavored over the years
during development of the adaptive optics system at Palomar.  In particular,
we have benefited substantially from conversations with T.~Hayward,
M.~Troy, R.~Dekany, and R.~Burress, and enjoyed the expert assistance at
the telescope of J.~Mueller and K.~Dunscombe.  J.~Carpenter also
participated in some of the data acquisition.  This publication makes
use of data products from the Two Micron All Sky Survey, which is a
joint project of the University of Massachusetts and the Infrared
Processing and Analysis Center/California Institute of Technology,
funded by the National Aeronautics and Space Administration and the
National Science Foundation.

\end{acknowledgements}
 
%\clearpage

%\clearpage  

\begin{deluxetable}{crcccccc}
\tabletypesize{\scriptsize}
\tablecaption{Near-infrared Point Sources in the Vicinity of Vega
\label{tab_phot}}
\tablehead{ \colhead{ID} & \colhead{Coordinates} & \colhead{$J$} &
  \colhead{$H$} & \colhead{$K_s$} & \colhead{Sep.\ from Vega } &
  \colhead{P.A.} & \colhead{Mass if assoc.\tablenotemark{\dag}} \\
  \colhead{} & \colhead{(J2000.0)} & \colhead{(mag)} & \colhead{(mag)} &
  \colhead{(mag)} & \colhead{(arcsec)} & \colhead{(degree)} & 
  \colhead{($M_{\rm Jup}$)} }
\startdata
1 & 18:36:58.19 +38:46:56.2 & $15.64\pm0.07$ & $14.78\pm0.05$ &
  $14.53\pm0.06$ & $22.26\pm0.03$ & $103.4\pm0.1$ & 13--35 \\  
2 & 18:36:58.08 +38:47:00.7 & $>18.5\pm0.1$ & $17.20\pm0.07$ & 
  $16.55\pm0.06$ & $22.33\pm0.03$ & $\phn91.8\pm0.1$ & \phn7--24 \\
3 & 18:36:58.70 +38:46:58.4 & $>19.3\pm0.1$ & $18.92\pm0.12$ & 
  $18.23\pm0.12$ & $27.70\pm0.03$ & $\phn96.0\pm0.1$ & \phn4--18 \\
4 & 18:36:59.35 +38:47:05.5 & $17.15\pm0.13$ & $16.25\pm0.14$ &
  $15.98\pm0.12$ & $29.41\pm0.05$ & $\phn86.2\pm0.1$ & \phn8--27 \\
5 & 18:36:59.39 +38:47:05.7 & $16.76\pm0.20$ & $16.29\pm0.16$ &
  $16.27\pm0.12$ & $29.93\pm0.05$ & $\phn85.8\pm0.1$ & \phn8--27 \\
6 & 18:36:55.36 +38:47:25.9 & \nodata & $17.43\pm0.07$ & \nodata &
  $27.05\pm0.05$ & $335.0\pm0.1$ & \phn6--22 \\
7 & 18:36:58.43 +38:46:37.9 & $17.12\pm0.12$ & $16.50\pm0.06$ &
  $16.20\pm0.07$ & $33.87\pm0.06$ & $133.8\pm0.1$ & \phn8--27 \\
8 & 18:36:58.40 +38:46:37.2 & $17.18\pm0.12$ & $16.48\pm0.09$ &
  $16.29\pm0.09$ & $34.11\pm0.06$ & $135.1\pm0.1$ & \phn8--27 \\
\enddata
\tablenotetext{\dag}{Minimum value interpolated from the \citet{bur01}
  models for 300~Myr; maximum value from the \citet{cha00} models for
  500~Myr.}
\end{deluxetable}

%\clearpage

\begin{figure}
\epsscale{1.1}
%\plottwo{fg1a.ps}{fg1b.ps}
\figcaption{(a) Composite $H$-band mosaic of the Vega region
  obtained with PALAO. Eight point sources are detected, 5 to the east,
  2 to the south-east, and 1 to the north of Vega.  The two close,
  eastern-most objects are just off the edge of the deep exposure of the
  Vega E field,  but have been pasted in from our shallower $JHK_s$
  images obtained for photometry purposes, to show their location.  
  Similarly, the double source to the south-east was discovered only in
  the follow-up shallow $JHK_s$ observations.  A
  bright ``ghost'' reflection of Vega is also visible in the north field.
  (b) The same image, with a smoothed ($\sigma=5$~pix) version of itself
  subtracted, to enhance faint sources in the wings of Vega's halo.
\label{fig_pharo}}
\end{figure}

\begin{figure}
\epsscale{0.7}
\plotone{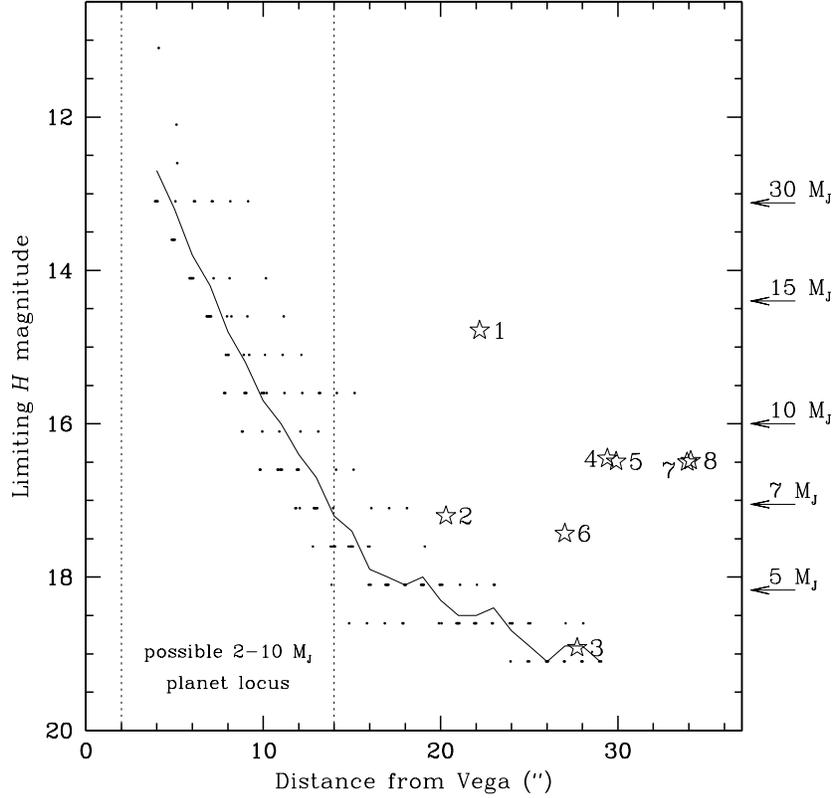}
\figcaption{$H$-band sensitivity of our deep images to faint objects
  as a function of radial distance from Vega, analyzed
  for the case of the east field. Solid points represent individual
  measurements of the limiting magnitude at different position angles
  and angular separations from Vega (a slight offset has been applied 
  between points along the abscissa for clarity).  The solid line 
  delineates the azimuthal
  average as a function of separation.  Numbered pentagrams indicate
  detected point sources. Horizontal arrows indicate the corresponding
  planetary mass at a given $H$ magnitude (for 300~Myr, Burrows,
  priv.\ comm.).  The area between the vertical dotted lines marks
  the locus of the inferred planet \citep{gor01,wil02}.  Thirteen per
  cent of the total area 
  imaged has twice the integration time and hence $\sim$0.4~mag better
  sensitivity, which is not accounted for in this analysis.  The
  limiting magnitude along the brightest ray at ${\rm P.A.}=50\arcdeg$
  (see Figure~\ref{fig_pharo}) is $\sim$1~mag poorer (as realized in the
  uppermost points in the graph) than along directions with no bright
  artifacts.  No limiting magnitudes are inferred for the
  $\sim$4$\arcsec$$\times$4$\arcsec$ area covered by the 
  ``ghost'' in the north field.
\label{fig_sens}}
\end{figure}

\end{document}